\documentclass{aa}
\usepackage{graphicx,natbib,ulem,txfonts,amssymb}
\input {epsf}
\def \xmm{{\it XMM-Newton\/}}
\def \cha{{\it Chandra\/}}
\def \h70 {$h_{70}$}
\begin{document}
\title{The late merging phase of a galaxy cluster : \\ XMM EPIC
  observations of A3266.} 
\author{J.L.~Sauvageot\inst{1}, E. Belsole\inst{2} and G.W. Pratt\inst{3}}
\offprints{jsauvageot@cea.fr}
\institute {$^1$C.E.A., DSM, DAPNIA, Service d'Astrophysique, C.E. Saclay,
  F91191, Gif sur Yvette Cedex, France\\
$^2$H.H. Wills Physics Laboratory, University of Bristol, Tyndall
Avenue, Bristol BS8 1TL, U.K.\\ 
$^3$Max-Planck-Institut f\"ur extraterrestrische Physik, 85748
Garching, Germany\\ 
} 
\date{Received date ; accepted date}
\abstract{We present a mosaic of five XMM-Newton observations of the
nearby ($z=0.0594$) merging galaxy cluster Abell 3266. We use the
spectro-imaging capabilities of \xmm\ to build precise (projected)
temperature, entropy, pressure and Fe abundance maps. The temperature
map exhibits a curved, large-scale hot region, associated with
elevated entropy levels, very similar to that foreseen in numerical
simulations. The pressure distribution is disturbed in the central
region but is remarkably regular on large scales. The Fe abundance map
indicates that metals are inhomogeneously distributed across the
cluster. Using simple physical calculations and comparison with
numerical simulations, we discuss in detail merging scenarios that 
can reconcile the observed gas density, temperature and entropy
structure, and the galaxy density distribution.
\keywords{Galaxies: clusters: individual: Abell 3266 - X-rays: galaxies:
  clusters - galaxies: intergalactic medium}
}
\titlerunning{An \xmm\ observation of A3266}
\authorrunning{J.-L. Sauvageot et al.}
\maketitle
 
\section{Introduction}
The last decade has seen increasing interest in the study of the
dynamics of galaxy clusters. Previously, cluster dynamics were almost
exclusively investigated by studying galaxy spatial and velocity
distributions using observations at optical wavelengths. 
However, projection effects and the inaccurate determination of
cluster galaxy membership can preclude firm conclusions about
sub-clustering and the dynamical state. X-ray observations are less
affected by projection effects and provide a powerful tool to
investigate substructure in clusters of galaxies. {\it Einstein} and
{\it ROSAT} data (combined with optical studies) have suggested  that
20 to 55 per cent of clusters show evidence of substructure in their gas
and/or galaxy distribution (e.g., \citealt{jf99},
\citealt{schueckeretal01}, \citealt{kolokotronisetal01}). The high
energy X-ray detectors aboard the ASCA satellite allowed for 
the first time the use of the temperature distribution of the
intra-cluster gas (ICM) as a quantitative tool for the determination
of the cluster 
dynamical state (e.g. \citealt{mark96}, \citealt{mark98},
\citealt{donnelly01}). These observations revealed
that many clusters which appear fairly smooth in their projected gas
and/or galaxy distributions can in fact show complex temperature
structure, indicative of dynamical activity.
 
The temperature distribution of the ICM can now be mapped
in exquisite detail with the instruments on board \xmm\ and \cha,
making it possible to explore the effects of the cluster formation history
on the thermodynamics of the gas, and to make detailed,
quantitative, dynamical models of individual merger events (e.g.,
\citealt{m02bs}; \citealt{bel04,bel05}; \citealt{hfb04};
\citealt{govoni04}). In this paper we use \xmm\ and \cha\ observations
to attempt to piece together the dynamical history of the nearby
($z=0.0594$) merging galaxy cluster Abell 3266 (A3266). 

A3266 has been extensively studied in the optical and the X-ray
wave-bands.  \citet{qrw96}, in an analysis of over 300 spectra over a
$1\fdg8\times 1\fdg8$ field of view, found a velocity
dispersion of 1400-1600 km s$^{-1}$ at the centre, decreasing to
700-800 km s$^{-1}$ towards the outskirts ($\sim 3 h_{50}^{-1}$
Mpc). They interpreted this velocity gradient as the effect of an old
merger which started some $4\times10^9h_{50}^{-1}$ years ago, with maximum
core collapse occurring $\sim 2 \times 10^9h_{50}^{-1}$ years ago. In
their scenario, the collision was in the southwest-northeast (SW-NE)
direction with a relative velocity between the two colliding objects of
$\simeq 1000$ km s$^{-1}$.

In X-ray, A3266 has been studied using  {\it ROSAT} \citep*{mme99},
ASCA \citep*{hdd00,mark98}, and BeppoSax \citep{dgm99}. All of these analyses
found signs of merger activity, including (i) elongated
central surface brightness along the SW-NE axis
\citep{mme99}; (ii) asymmetric temperature features and a decreasing
temperature profile from $\sim10$ in the centre to $\sim5$ keV
\citep{mark98,dgm99,hdd00} at a distance of $15 \arcmin~ (\sim 1
h_{70}^{-1}~{\rm Mpc}$); and (iii) asymmetric variations in the
metallicity distribution \citep{dgm99}. The latter authors found a
relatively low average metal abundance of $0.17\pm0.02~Z/Z_{\odot}$. 
 
\citet{rf00} developed a 3D numerical model using the optical data from
\citet{qrw96} in combination with constraints from {\it ROSAT} and
ASCA X-ray observations. They suggested that A3266 is the result of an
old ($\simeq 3$ billion years ago), off-axis collision of two clusters
with a mass ratio of $\sim 1:2$. In their picture, the less massive
sub-cluster crossed the dominant cluster in the SW-NE direction,
passing the western 
side of the dominant cluster while moving into the plane of the sky at
$\sim45^{\circ}$.  Interestingly, they also predicted a large amount
of angular momentum transfer into the ICM from the two colliding
clusters. This momentum transfer should be measurable with X-ray
bolometers, but unfortunately cannot be verified with the CCDs on
board \xmm\ (or {\it Chandra}), as approximately $\simeq10$ times
better spectral resolution is needed (see also \citealt{rf00}). 

Recently, \citet{ht02}  presented a {\it Chandra} observation of
A3266. This was obtained with the ACIS-I detector, and the centre of
the cluster was centred on chip 1. Unfortunately the roll-angle caused
the whole north-eastern elongation of the cluster to be missed,
although the \cha\ data allow a good examination of the central
4\arcmin. In agreement with previous work, \citeauthor{ht02} found a
decreasing temperature profile and observed local temperature
variations on scales of 0.25\arcmin\ as defined by the temperature map. 
They also found an 
enhancement in a relatively flat abundance profile that they interpreted
as the result of metal deposition from a higher
metallicity sub-cluster into the main cluster as a consequence of the
merging event. The overall picture resulting from these \cha\ data,
together with their re-investigation of the Quintana et al. galaxy
distribution was of a relatively minor merging event occurring in the
plane of the sky.   

Here we present a study which, by taking advantage of the high
sensitivity and large field of view provided by \xmm, surpasses
previous high quality X-ray observations.  We combine the 5
overlapping publically-available \xmm\ pointings into a mosaic
covering $\sim 20\arcmin$, and use this to investigate the merger
scenario in A3266. Our conclusions are the result of the
interpretation of precise temperature, pressure and entropy maps, in
combination with archival \cha\ data and optical observations.

Throughout this paper we use a cosmology with $h = H_{\rm 0}/100 =
0.7$, $\Omega_{\rm M}$=0.3 and $\Omega_{\Lambda}$=0.7. In this
cosmology and at the cluster redshift, 1\arcmin~ corresponds to 68
kpc. If not otherwise stated, errors are quoted at 1$\sigma$ for one
interesting parameter and all abundances are given relative to solar
using the tables of \citet{gs98}.

\section{Data preparation }
\subsection{\xmm\ observations and mosaic construction}
\begin{figure}
\centering
\includegraphics[scale=0.45,angle=0,keepaspectratio,width=\columnwidth]{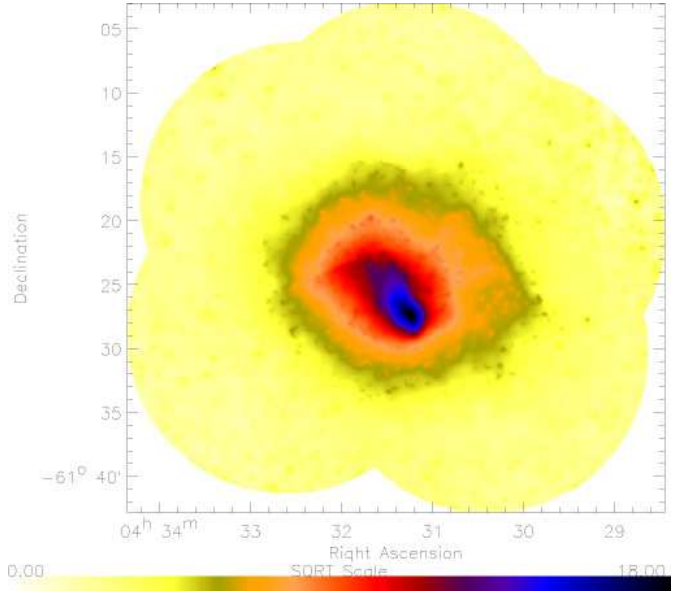}
\caption{Low energy (0.5-2.5 keV) wavelet reconstructed EMOS X-ray image of
A3266. The image clearly shows the strong change of isophotal
orientation with distance from the centre, the compression of the
isophotes toward the SW in the central region, and the bright
off-centre core.}
\label{fig:expmap_LE}
\end{figure}

The EPIC observations were retrieved from the \xmm\ Science Archive
(XSA) database. A total of six $\sim 30$ ks pointings cover the entire X-ray
extent of the cluster in the sky. Observation details are given in
Table~\ref{tab:obs}.  As a background we used the
blank-sky background described in \citet[][hereafter the DL event
list]{dlbkg}.

\begin{table}
\begin{center}
\caption{Journal of observations. The exposure time given corresponds
to that remaining after cleaning for soft proton flares. Observation
0105262001 was not used since it was heavily comtaminated by flares.} 
\label{tab:obs}
\begin{tabular}{c c c c c }
\hline
\multicolumn{1}{l}{Obs-ID} & \multicolumn{1}{l}{Date} &
\multicolumn{3}{c}{Clean exposure (s)} \\ 
\multicolumn{1}{l}{ } & \multicolumn{1}{l}{ } &
\multicolumn{1}{l}{EMOS1} & \multicolumn{1}{l}{EMOS2} &
\multicolumn{1}{l}{EPN} \\ 
\hline
0105260701 & 2000-10-01 & 21385 & 21393 & 17498 \\ 
0105260801 & 2000-10-11 & 20743 & 20743 & 16849 \\
0105260901 & 2000-10-09 & 24942 & 24942 & 21049 \\ 
0105261001 & 2000-09-27 &  5756 &  5691 &  6214 \\  
0105261101 & 2000-09-25 & 13892 & 13891 &  9998 \\
\hline
\end{tabular}
\end{center}
\end{table}               

We screened the event lists for flare-induced background contamination
by generating light curves in a high energy band, where counts were
grouped in 100 second bins. Background files were filtered with the
same criterion adopted for the source events. This consists of 
rejecting those time intervals with more than 18 cnts/100s in the
10-12 keV energy band for MOS cameras, and 22 cnts/100 s in the 12-14
keV energy band for the pn, respectively. One of the pointings
  was heavily contaminated by proton flares, leaving us with
five useful pointings.  We then built a mosaic
event list by merging all the events from each pointing. To build the
background merged event list, we rotated the original DL event lists
to match the sky coordinates of each of the five pointings. It was
then possible to merge these five background event lists to create a
background mosaic event list.

To take into account vignetting of the telescope, we used the photon
weighting technique computed with the task {\sc evigweight} in the
Science Analysis System (SAS). The overlapping regions of the mosaic
were accounted for by dividing the weight associated with each event
by the total exposure time. This allows us to use the merged event
list as if it was a single pointing. 
The total exposure time per camera achieves $\sim 80$ ks in the
central part of the cluster.

\subsection{Chandra observation}

A 30 ks \cha\ observation of A3266 was performed in July 2000 using
the ACIS-I detector in VFAINT mode (ID 899). We created a new level-2
event file using CIAO 3.1, following the recommended procedures
described in the \cha\ data analysis\footnote{ {\tt
http://asc.harvard.edu/ciao/threads/createL2/}}. In particular we
filtered for periods of high background (negligible for this data
set), corrected for gain variation, 
and applied an improved astrometry to the data.
Our main interest in using these data is to map the low-energy X-ray
emission from the central regions of the cluster at the best available
spatial resolution, in order to compare these results directly with
\xmm\ data. We thus detected and excluded point sources and filled
these regions with Poisson noise using the CIAO tasks {\it wavdetect}
and {\it dmfilth}, respectively. We then used wavelet filtering to
build a smoothed X-ray image.  

\section{X-ray morphology} 

\begin{figure}
\vspace{-0.25cm}
\includegraphics[scale=0.45,angle=0,keepaspectratio,width=\columnwidth]{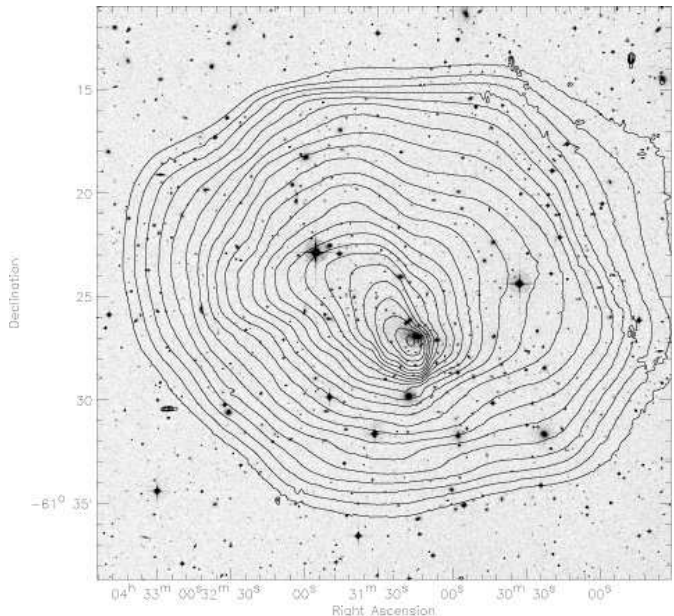}
\caption{Digital Sky Survey image of A3266 with low energy
    (0.2-2.5 keV) XMM EPIC contours superimposed.} 
\label{fig:DSS_LE}
\end{figure}

Figure \ref{fig:expmap_LE} shows the vignetting
corrected, \xmm\ 0.5-2.5 keV
energy band image of the wavelet reconstructed 
 mosaic in square root scale. The X-ray peak is found at
R.A. = $04^h31^m14^s1$, Dec. = $-61^{\circ}27\arcmin26\farcs5$
(J2000), in agreement with  the position of the X-ray peak
from \cha\ R.A. = $04^h31^m13^s5$, Dec. = $-61^{\circ}27\arcmin12\farcs0$
(J2000), taking into account that \xmm\ does not resolve emission from
the central galaxy. Figure~\ref{fig:DSS_LE} shows a Digital Sky 
Survey (DSS) image with the EPIC low energy X-ray contours superimposed. 

While the large scale morphology is fairly regular, in the inner 
region ($\sim 450 h_{70}^{-1}$ kpc) the isophotes are clearly
elliptical, and the bright inner core is unequivocally off-centre with
respect to the larger scale gas distribution. We observe a significant
compression of the isophotes from the peak of emission toward the
SW. In the opposite NE direction, we observe an
elongation of the surface brightness in a shape resembling a comet
tail. Limited statistics in the \cha\ data do not allow us to achieve
the 1\arcsec\ resolution limit, but the final wavelet reconstructed
\cha~ low energy image confirms the features found with the \xmm\
data (see also \citealt{ht02}).

\section{X-ray cartography}

It is well known that the temperature, entropy and pressure
distributions can offer key insights towards an understanding of the
physical conditions in merging clusters. In this Section, we use \xmm\
data to derive precise two-dimensional maps of these quantities, plus
the Fe abundance distribution, in A3266.

\subsection{Temperature map}

\subsubsection{The Voronoi Tessellation (VT) method}

Here we apply, for the first time, the Cappellari \& Copin
(2003; hereafter CC03) Voronoi tesselation algorithm (hereafter VT) to
build the \xmm\ EPIC temperature map of A3266. We first extracted
source and background images in the 0.5-7.5 keV energy band. The
chosen energy range optimises the cluster signal over the particle background
in the temperature range of the cluster. These two images were then
used to estimate the Signal-to-Noise (S/N) of each pixel.

We were unable to
use the CC03 algorithm in a single step, since X-ray events are
distributed following Poisson statistics and many of the pixels have a
low S/N. Our implementation of the algorithm thus involves two
steps. We first selected only those pixels with a sufficiently high
S/N (i.e. $(S-N)/N \ge 1.05$) and used the CC03 algorithm to bin these 
pixels into meta-pixels with a S/N $\simeq130$ (our final goal
for the temperature map). Since these meta-pixels were obtained from a
high S/N subset, they are not generated from a continuous set of
pixels. The second step consists of assigning all of the so-far
unbinned pixels to their closest meta-pixel. Obviously, the addition
of these lower S/N pixels adds scatter to the S/N of the final
distribution of meta-pixels. However, the resulting distribution of
convex meta-pixel cells covers the whole image without significantly
degrading the S/N.

\begin{figure}
\includegraphics[scale=1.,angle=0,keepaspectratio,width=\columnwidth]{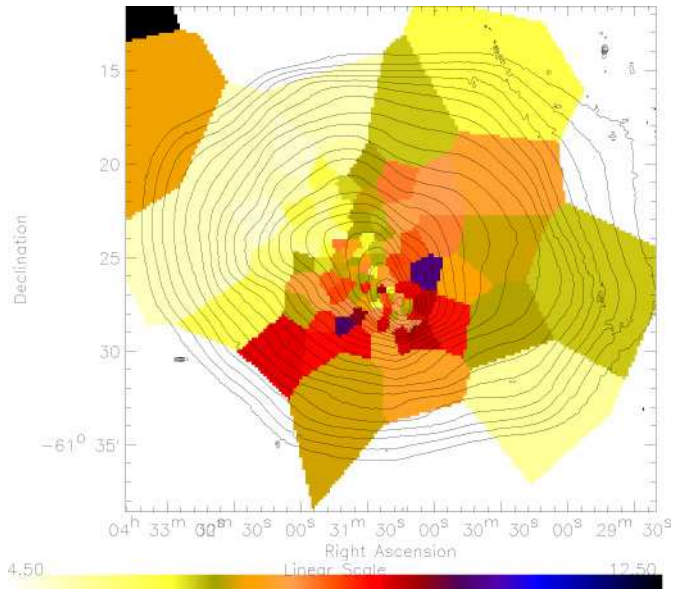}
\caption{{\footnotesize Temperature map obtained using the VT
    algorithm. The contours are from the wavelet reconstructed 0.5-2.5
      keV image with point sources excised. The hot (dark blue/red)
      region is very 
  likely a shock wave propagating toward the outskirts of the
  cluster. }}
\label{fig:Tmap}
\end{figure}

\begin{figure}
\includegraphics[scale=1.,angle=0,keepaspectratio,width=\columnwidth]{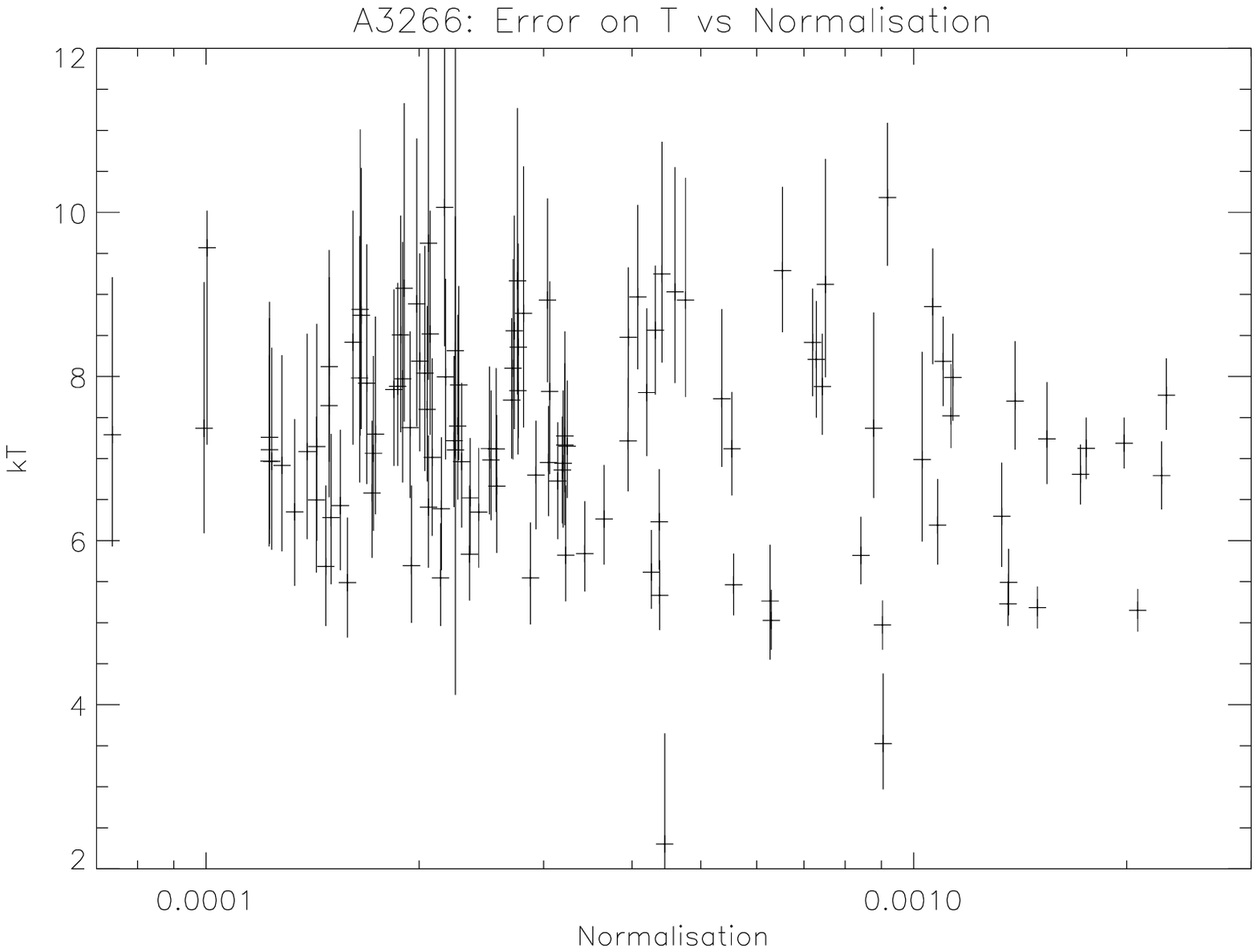}
\includegraphics[scale=1.,angle=0,keepaspectratio,width=\columnwidth]{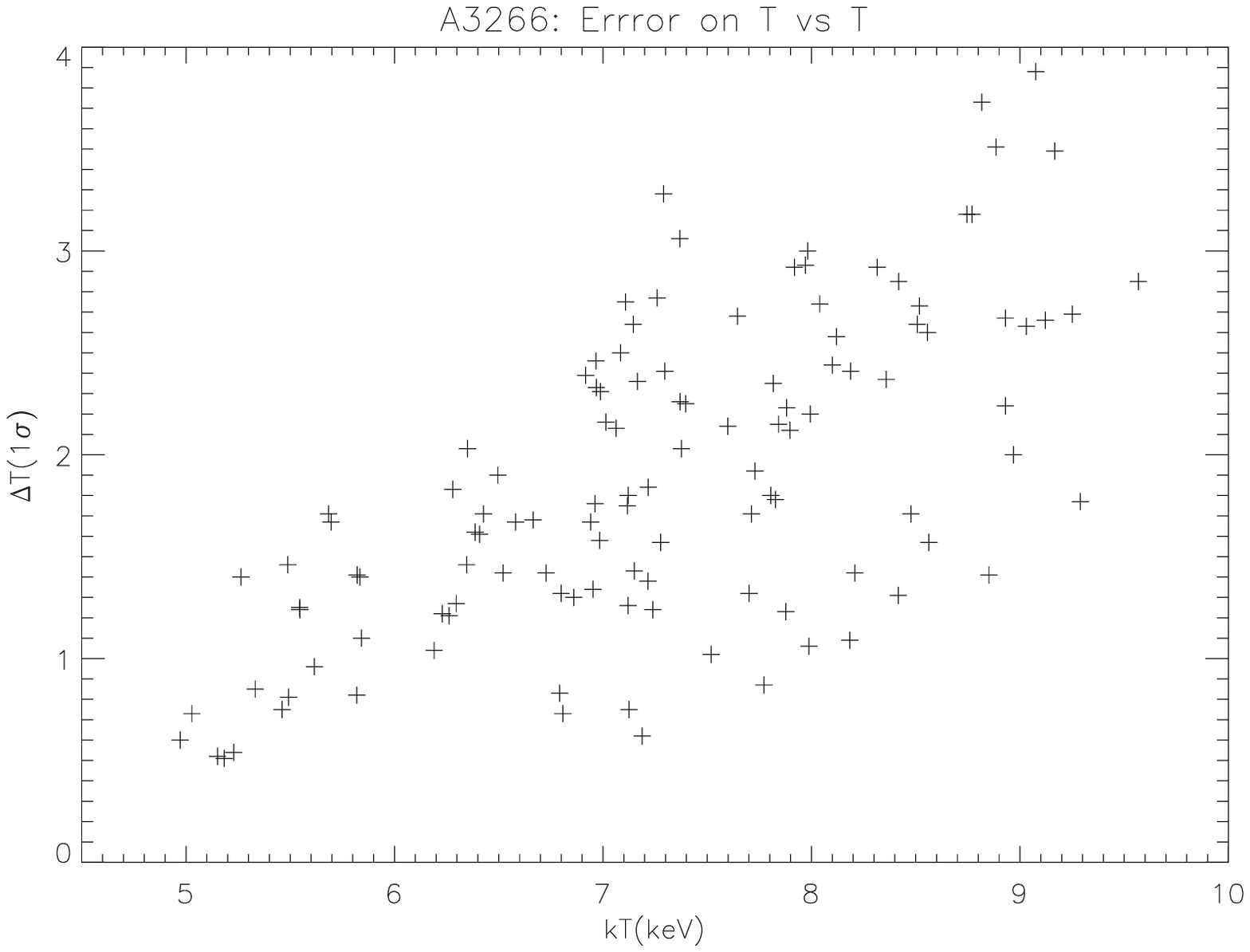}
\caption{Error on the temperature determination. Mean error bars are
  k$T^{+1.2}_{-0.9}$ keV.
In the top panel we plot the error on $T$ versus the Emission Measure in
each meta-pixel. 
In the bottom panel we plot the difference between the two $1\sigma$
limits versus the fit temperature. The dispersion at a given
temperature shows the limit of the approach.} 
\label{fig:errT}
\end{figure}

Application of this technique to the mosaic event list of A3266
results in 138 cells. We fitted the spectrum of each cell with an
absorbed {\sc mekal} model using XSPEC v11.2. The absorption was fixed
to the Galactic value ($N_{\rm H} = 1.6\times10^{20}$ cm$^{-2}$;
\citealt{nh}), and the abundances were fixed to 0.2 $Z/Z_{\odot}$.
This abundance value was
obtained by fitting a global spectrum, extracted in a circle of
10\arcmin\ and excluding point sources, with a two-temperature {\sc
mekal} model with abundances tied together. Our best fitting global
abundance is in reasonable agreement with the
value found by \citet{ht02} in the central region mapped by \cha\ (see
also \citealt{dgm99}). We note that a single temperature model is an
acceptable fit to all 138 cells.

The VT temperature map, shown in Fig.
\ref{fig:Tmap}, was obtained by filling each cell area with the
best fitting temperature value. The 1$\sigma$ errors associated with
the temperature of each cell are plotted as a function of emission
measure on the upper panel of  Fig.~\ref{fig:errT}. The mean
$1\sigma$ error is  k$T^{+1.15}_{-0.86}$. In the lower panel of
Fig.~\ref{fig:errT} we plot the difference between the upper and lower
$1\sigma$ errors versus the best-fitting temperature. The observed
dispersion at a given temperature shows the limit of the approach. 
Since the photon statistics are a function of energy, the higher the
temperature, the larger the associated statistical error. 

\subsubsection{The wavelet spectral mapping method}

Like any adaptive binning algorithm, the VT technique samples the sky
plane as a set of independent meta-pixels within which the gas
temperature can be estimated. While this approach has enabled us to
check, in a straightforward manner, the goodness of fit in each cell,
it cannot achieve an accurate map of the complex structure of the gas
temperature. A more complete exploration of the scale-space can be
provided using the wavelet transform. We have therefore built another
temperature map using a specially-developed wavelet spectral mapping
algorithm. In this technique, the gas temperature is first estimated
in square resolution elements at different scales, allowing
characterisation of the spatial variations as Haar wavelet
coefficients. The temperature map is obtained from structures selected
as significant in the scale-space using a regularised reconstruction
process. Further details on the method can be found in \citet{hb04}.

Achieving a spatially homogeneous temperature estimation across the
field of view, whatever the local statistics, effective area or
background contribution, required some adaptations to the algorithm
described in \citet[][see also \citealt{bel04,bel05}]{hb04}. For this
mosaic, we associated a global 
emission model $F(T)$ to the overall data set, and maximised the
log-likelihood function $\log L(F(T))=\sum_{i=1}^N \log F_i(T)$, in
fitting $F(T)$ to the data. $F(T)$ is composed of a linear combination
of the expected source and 
background contributions to each pointing of the mosaic
observation. Calling $S(T,e)$ and $B(e)$ the normalised source and
background spectra associated with the pointing $p$, $F(T)$ can be
expressed as a function of the local exposure time $t_p$, and
effective area $AE_p$, as:

\begin{eqnarray}
F(T,k,l,e) = & \sum_{p=0}^{npoint-1} t_p(k,l) * AE_p(k,l,e)  & \nonumber \\
             &  * ns(k,l) * S(T,e) + nb_p(k,l) * B(e) &
\end{eqnarray}
where $e$ is the incident photon energy, $k$ and $l$ are the
meta-pixel coordinates at the current spatial scale, and $n$s and
$nb_p$ are the local emissivity terms associated with $S$ and $B$,
respectively. The source model
$S(T,e)$ is an absorbed {\sc mekal} plasma emission spectrum (with
$N_{\rm H}$ and abundances fixed as described above). The background
model $B(e)$ is a multicomponent spectrum which takes into account
contributions from the particle background and the cosmic X-ray
background (see \citealt{hb04} for details).

\begin{figure}
\includegraphics[scale=1.,angle=0,keepaspectratio,width=\columnwidth]{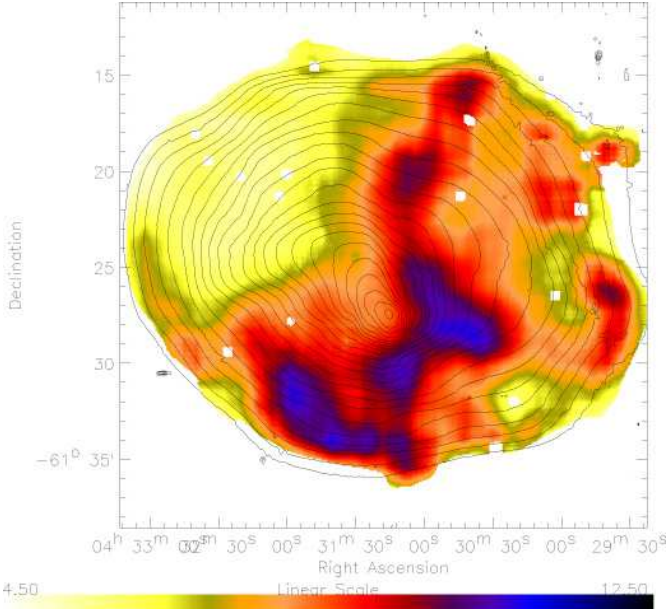}
\caption{{\footnotesize Temperature map built using the wavelet spectral
  spectral mapping code of \citet{hb04}. The probable shock wave region is
  again very clearly seen. White squares are excised point sources.}}
\label{fig:TmapXR}
\end{figure}

The resulting temperature map, shown in Fig. \ref{fig:TmapXR}, was
obtained from a wavelet analysis performed on 
7 scales, corresponding to structures of typical sizes
ranging from 10\arcsec~ to 5\arcmin. All the structures were detected
according to a 3 $\sigma$ significance criterion. As expected, the
spatial resolution is better near the cluster core, where the
emissivity is the highest, than in the external regions, which have
lower photon statistics.
Point-by-point comparison is difficult because of the
large cell size at the edges of the VT map. Nevertheless, comparing
Figs.~\ref{fig:Tmap} and~\ref{fig:TmapXR} shows that the temperature
  maps are consistent within an uncertainty of $\pm 0.5$ keV.

\subsubsection{Description of the temperature structure}

A3266 exhibits temperature structure at several different spatial
scales (Fig. \ref{fig:TmapXR}). First of all, we can identify a region
of enhanced temperature which follows closely the compression of the
low energy isophotes SW of the X-ray emission peak. This region
extends out to 12\arcmin\ ($\sim 0.8 h_{70}^{-1}$ Mpc) from the
cluster centre, with two `arms' pointing toward the north and the SE.
The temperature achieves 10 keV in this region, while the surroundings
are found to be at $\sim$6-7 keV.  Secondly, it is clear that the
inner core of the cluster, encompassing the X-ray emission peak, is
cooler (k$T\sim$8 keV) than the region directly to the SW. Lastly,
going from the X-ray emission peak toward the NE, we observe a smooth
decline from k$T$=8 keV, in the centre, to k$T\sim$ 5 keV, at the NE
edge of the map.
These XMM data show considerably more temperature detail than any previous
X-ray observations of this cluster. The overall temperature structure
is in good agreement with the ASCA analysis (\citealt{mark98};
\citealt{hdd00}) once averaged over the ASCA PSF. 
The \cha\ temperature map presented in
\citet{ht02} has rather poor spatial resolution as the data suffer from
the limited photon statistics. However, once averaged to the spatial
resolution of the \cha\ temperature map, the \xmm\ map
looks similar.

We note that the main temperature structure is very similar to the
predictions of numerical simulations of on-axis mergers between
clusters in the compact/central phase, when a moderately supersonic
shock develops toward the outskirts of the newly formed cluster (e.g.,
\citealt{rbl96}, \citealt{rs01}), as discussed below in
Sect.~\ref{sec:disc}. 

\subsection{Entropy and pressure maps}

\begin{figure*}[]
\begin{tabular}{c c c}
\includegraphics[scale=1.,angle=0,keepaspectratio,width=0.33\textwidth]{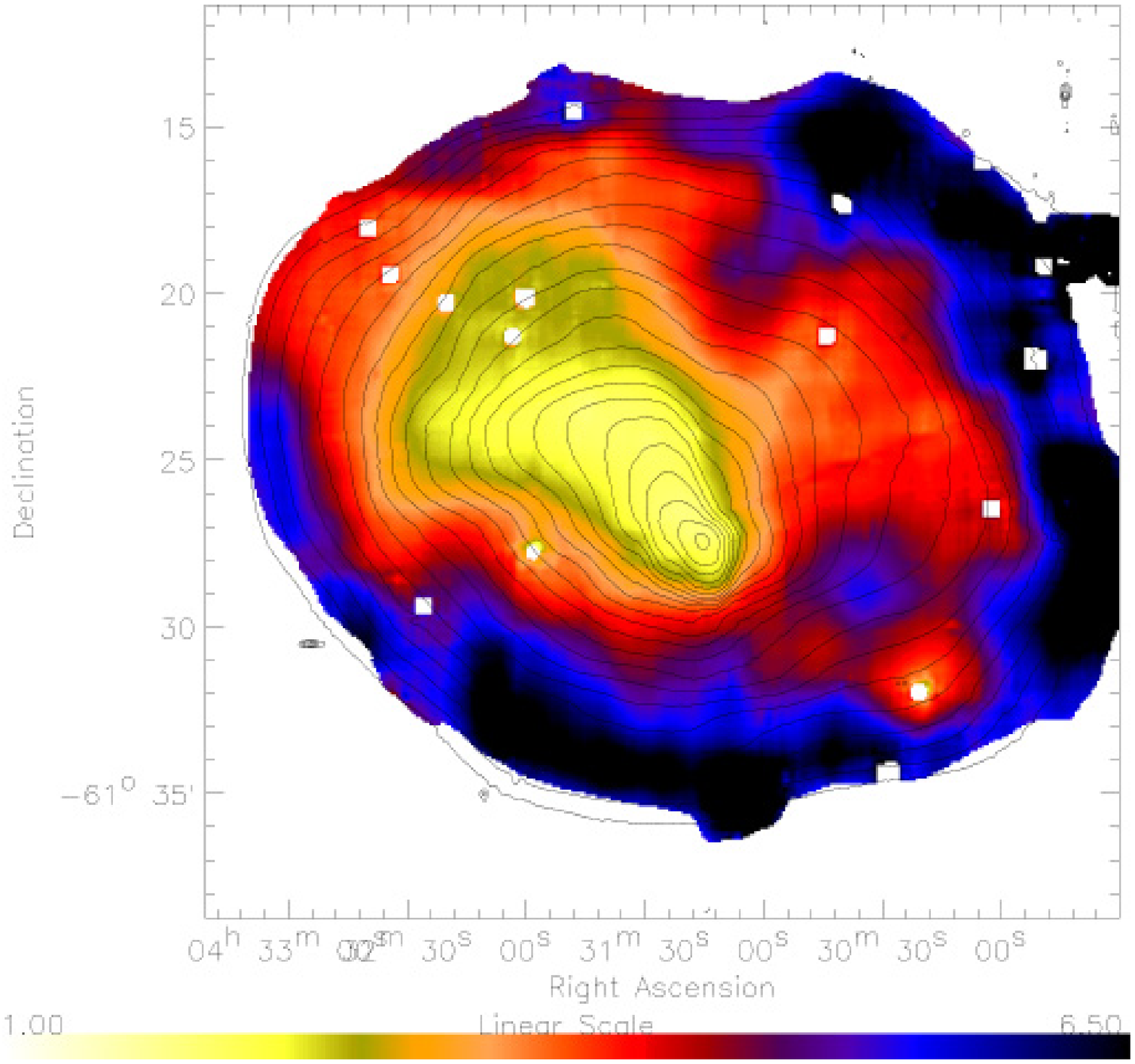}
\includegraphics[scale=1.,angle=0,keepaspectratio,width=0.33\textwidth]{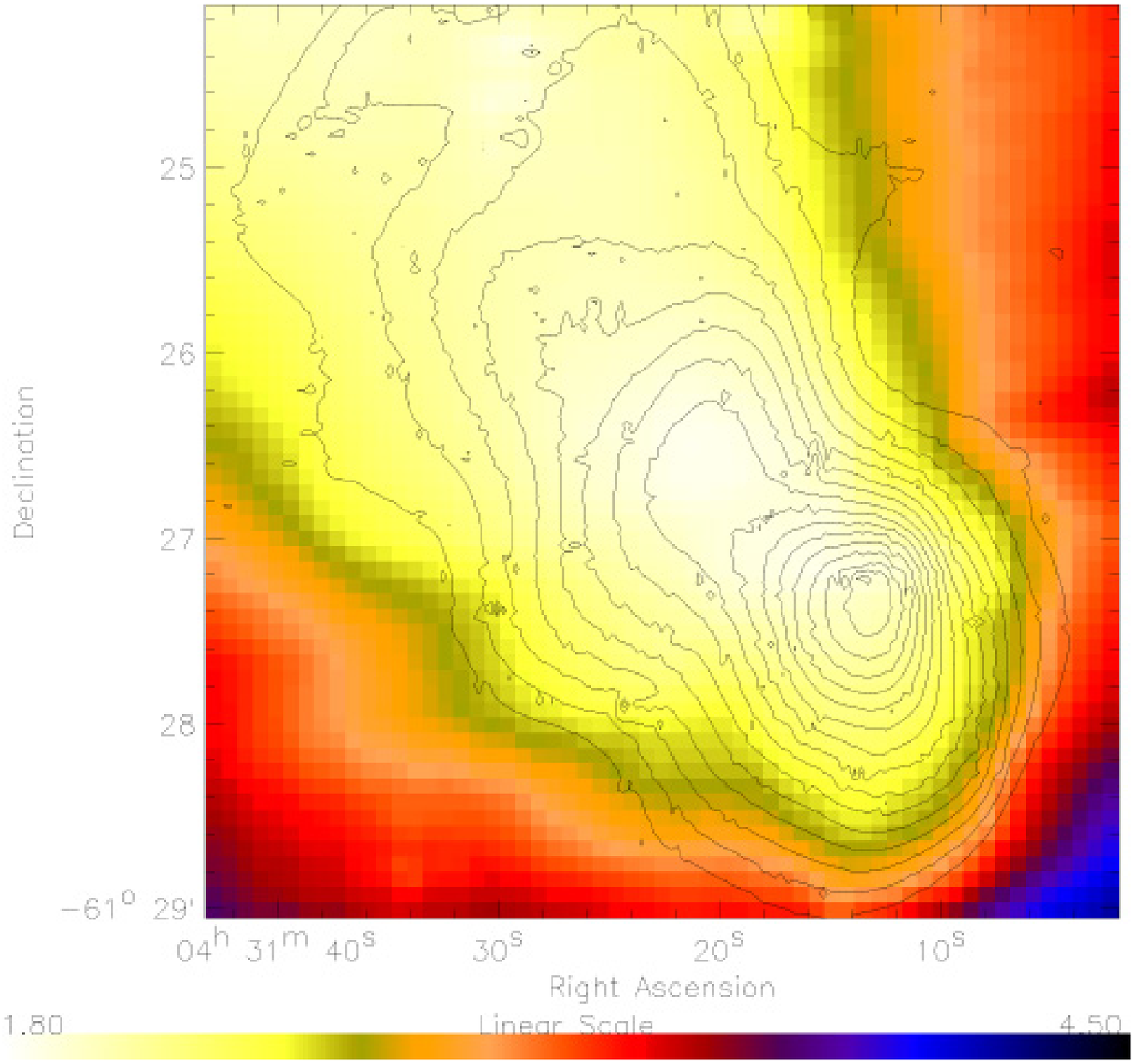}
\includegraphics[scale=1.,angle=0,keepaspectratio,width=0.33\textwidth]{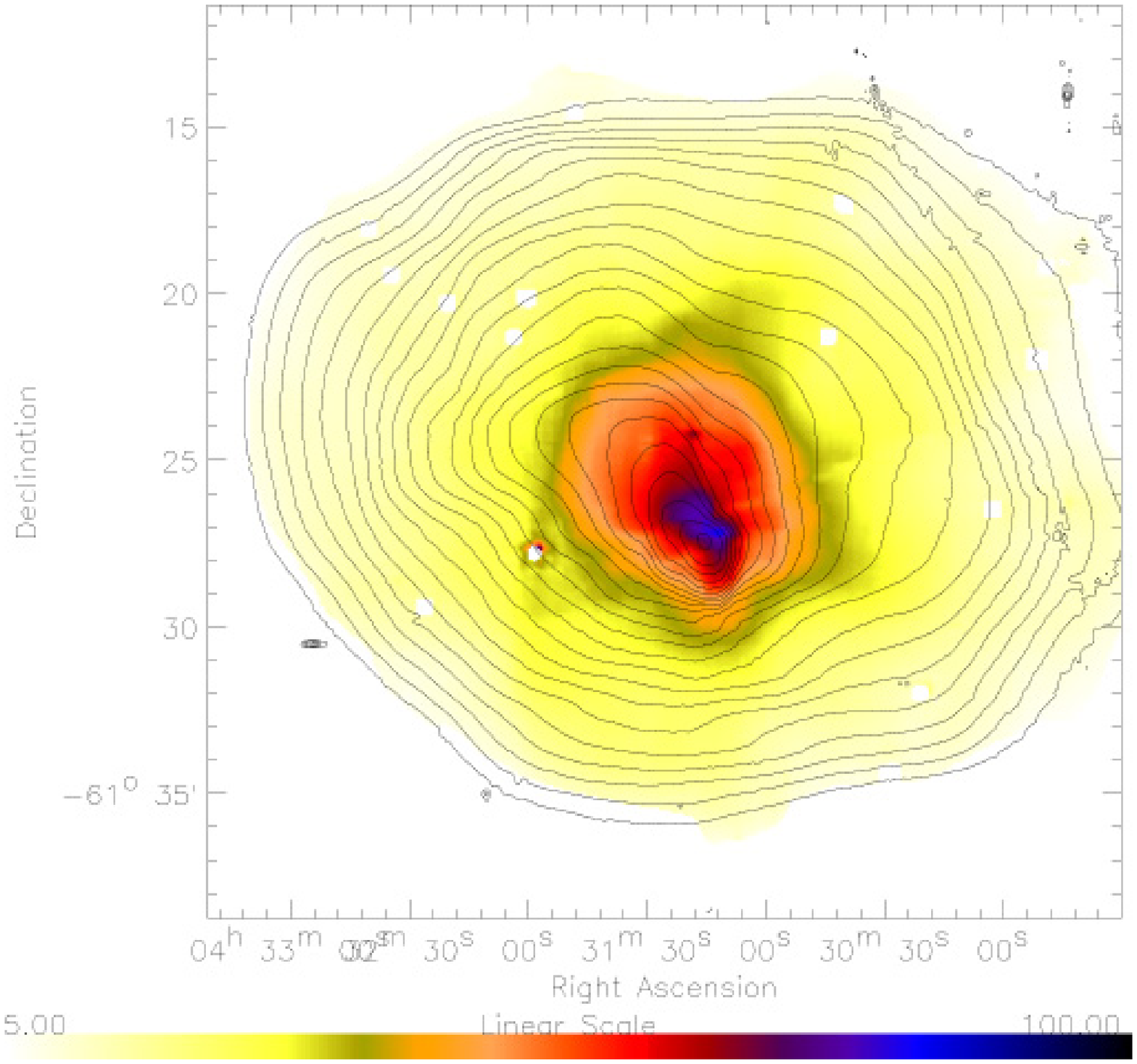}
\end{tabular}
\caption{{\bf Left}: Projected entropy map defined as $S\propto T
/I^{1/3}$ (arbitrary units); {\bf Middle}: Zoomed image of the
projected entropy distribution in the cluster core, with contours
of the 0.5-2.5 keV \cha\ image overlaid. Note how closely the contours
of the low energy X-ray emission follow the entropy gradient; {\bf
Right:} Projected pressure map defined as $P\propto T *\sqrt{I}$
(arbitrary units). The contours are from the wavelet reconstructed
0.5-2.5 keV EPIC image with point sources excised. The regularity of the
pressure distribution is noticeable. }
\label{fig:PS}
\end{figure*}
We obtained ``pseudo'' (projected) entropy ($S$) and pressure ($P$)
maps from our emissivity and temperature distributions using the
following definitions : $S\propto T / I^{1/3}$ and $P\propto T
*\sqrt{I}$, where $I$ is the 0.2-2.5 keV X-ray image. 
These maps are shown in Fig. \ref{fig:PS}.  The pseudo-entropy map
(left-hand panel of Fig.~\ref{fig:PS})
displays a very elongated dip, which is associated with the comet tail
shaped surface brightness distribution in the centre of the cluster. In
the middle panel of Fig.~\ref{fig:PS}, we show a zoomed view of the
central entropy 
with low energy (0.5-2.5 keV) \cha\ contours overlaid. The contours of
the low energy emission follow very closely the strong gradients in
the \xmm\ entropy map. Interestingly, the pressure (right-hand panel
of  Fig.~\ref{fig:PS}) exhibits much more
axial symmetry than the entropy, temperature or emission measure maps. 

\subsection{Core Fe abundance distribution}

The 80 ks exposure achieved in the central region with the \xmm\ mosaic 
allows us to investigate the spatial distribution of
metallicity. We do this by generating an Fe equivalent width (EQW) map. We
extracted three images in the following energy bands: 6000-6453 eV  (Fe
K line), 5667-5995 eV (low energy continuum), and 6460-6913 eV (high
energy continuum). These images were smoothed adaptively using the
same template\footnote{The template was built on the basis of count
  statistics as defined by the radial profile of the 6-6.45 keV
  image. With this type of template, the smoothing box increases
  linearly from the centre to the outer regions, following the radial
  count statistics. This method of template definition avoids the
  formation of spurious structure sometimes observed in adaptive
  smoothing.}. We estimated the continuum under the line by
interpolation between the two continuum images. The EQW image
was then obtained simply by dividing the Fe K image by the interpolated
continuum image. The EQW map is sensitive to temperature variations
  through the Fe emissivity. We have used the table of Fe line total
  equivalent widths from \citet{ra85}, together with our temperature
  map, to convert the EQW map to a spatially resolved abundance
  distribution.

\begin{figure}
\vspace{-0.25cm}
\includegraphics[scale=1.,angle=0,keepaspectratio,width=\columnwidth]{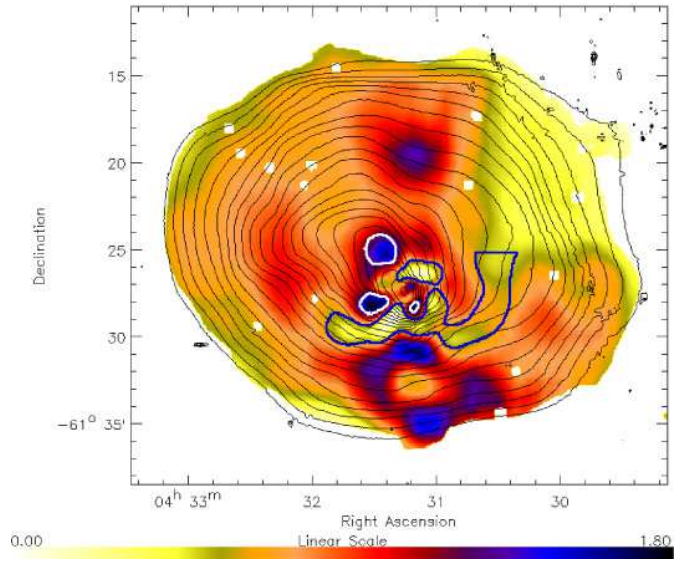}
\caption{{\footnotesize Fe K abundance map of the cluster. The low-abundance
    (light-yellow) ``path'' is confirmed by a spectral fit of the region
    enclosed by the blue line ($Z/Z_\odot=0.17\pm0.05$).
The high abundance regions enclosed in white have  $Z/Z_\odot = 0.23 \pm0.05$, when
fitted together.}}
\label{fig:EQW}
\end{figure}

The resulting Fe abundance distribution (Fig.~\ref{fig:EQW}) exhibits a
remarkable low abundance `path', which
appears to extend between a more uniform region of higher abundance.
We extracted a spectrum of this ``path'', following the region
outlined in blue in Fig. \ref{fig:EQW}. The best fitting values of
$Z/Z_\odot=0.17\pm0.05$ and k$T=7.5\pm0.3$ keV
confirm the low abundance of this region. Another spectrum was
extracted on the basis of the small-scale higher metallicity areas
(defined by white circles in Fig.~\ref{fig:EQW}). This spectrum is
best fitted with a metallicity of $Z/Z_\odot=0.23\pm0.05$ and
k$T=6.36\pm0.3$ keV. These direct spectral measurements thus confirm
the validity of the method we have used. The Fe map appears to
indicate that the 
gas in the cluster core is not yet well mixed.

\section{Optical observations}

\subsection{Galaxy density map}
\label{sec:galden}

Galaxy relaxation after a cluster merger is commonly thought to occur
at an intermediate speed between that of the gas and that of the dark
matter.
Study of the galaxy distribution and dynamics thus gives further
insight into merger kinematics than that achievable through study of
the gas alone.  We have therefore built a projected density map with
the algorithm developed by \citet{slez05}, using the galaxies in the
APM catalogue \citep{mad90}. Use of such a map has the great advantage
of good statistics, thereby allowing detection of subtle variations of
the density field. At the same time, projection effects cannot be
excluded without a systematic redshift confirmation.

We used galaxies with $R\le 19$ (corresponding to $L_*+2.5$) to build
the galaxy density map. This magnitude was chosen as it allowed
sufficient sampling of the cluster while limiting background
contamination. In addition, at fainter magnitudes, star-galaxy
separation becomes critical in the APM catalogue. 

\begin{figure}[]
\begin{centering}
\includegraphics[scale=0.55,angle=0,keepaspectratio]{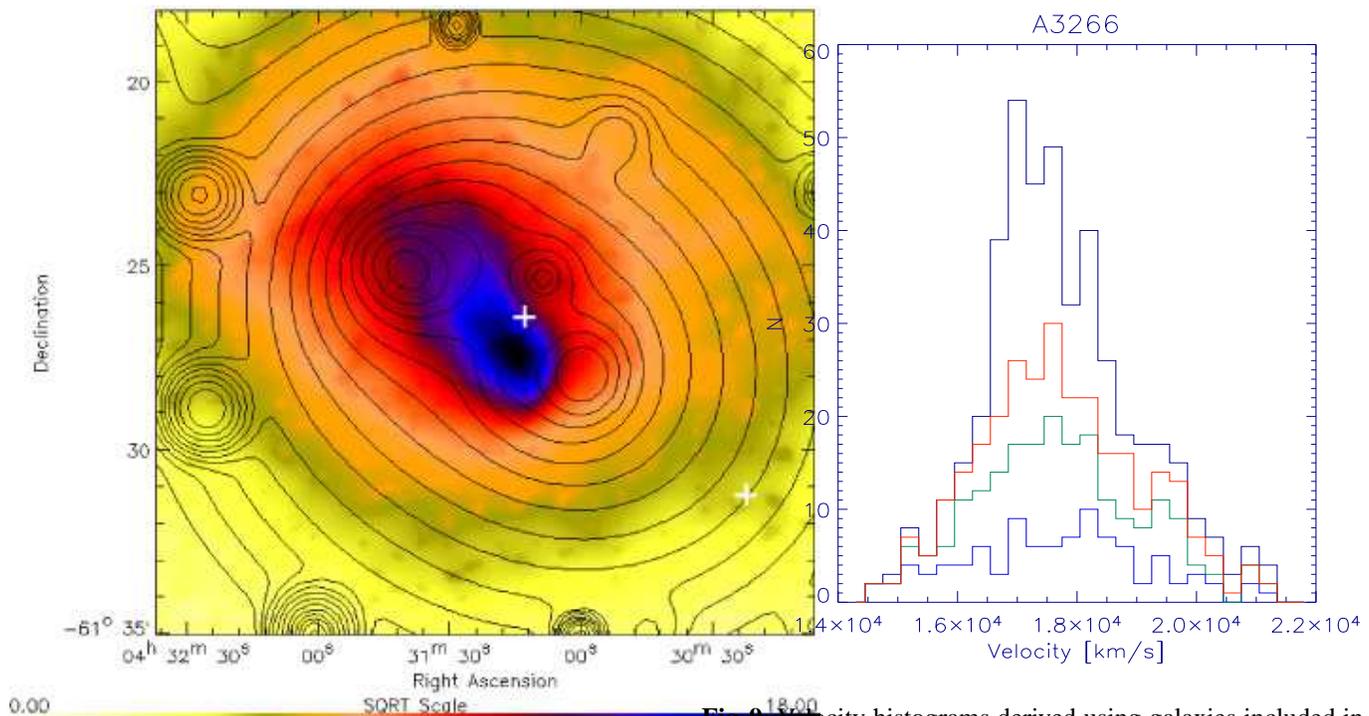}
\end{centering}
\caption{{\footnotesize Projected galaxy density map. The colour image
is the low energy X-ray image shown in Fig. \ref{fig:expmap_LE}. The
smoothed galaxy 
density (galaxies with $R \le 19$ from the APM catalogue) is displayed as
logarithmically-spaced contours. One clearly see the two density peaks
and the X-rays maximum lying in between. The central cross is located
at the position of the brightest cluster galaxy (BCG). The southwest cross
shows the position of the WAT galaxy.}}
\label{fig:DG}
\end{figure}

Using a wavelet
algorithm to build the galaxy density map allows a more careful
restoration of significant structure than the classical Dressler
method, and yields a reconstructed image with optimised spatial
resolution following count statistics. The final projected galaxy
density map is shown in Fig.~\ref{fig:DG}, where we observe two
strongly significant peaks to the NW and SE of the X-ray emission
maximum. The separation between the peaks is $\sim5.2$\arcmin~($\simeq
360h_{70}^{-1}$ kpc). The brightest galaxy is located at R.A. =
$04^h31^m13^s$ Dec.= $-61^{\circ}27\arcmin13\arcsec$, to the NE
extreme of the SW peak. This galaxy is a dumb-bell galaxy
\citep{hdd00}, with components oriented in the NE-SW direction. Other
peaks appearing in the image arise because of the wavelet method we
used. A chance projection of a small number of galaxies (typically
three or four) may appear as a highly significant local maximum at
high spatial frequency, and it is reconstructed as such in the final
image, even though we set the threshold for structures to be
significant to 3$\sigma$. We have visually checked the structures by
overplotting the projected galaxy density map on the APM
catalogue. All small-scale, high significance structures in
Fig.~\ref{fig:DG} come from projections of four or less galaxies.

We have compared our projected galaxy density map to previous
analyses. \citet{hdd00} and \citet{fqw00} find only a single central
peak, although both maps show a distinct elongation in the
NE-SW direction. There is a weak secondary peak at $\sim
4.8\arcmin$ ($\sim 325$ kpc) from the main galaxy concentration in
\citeauthor{fqw00}\hspace{-1mm}'s map. In our galaxy density map, the
second peak 
is found at a similar projected distance, but is more significant.
This discrepancy can be explained by the different data sets used. In
\citet{fqw00}, only galaxies belonging to the cluster with a measured
redshift were used, resulting in only 12 objects in the central
$5'\times5'$ region in \citeauthor{fqw00}, against 42 objects used for
calculating the map shown in Fig.~\ref{fig:DG}. The NE maximum
was very poorly sampled and the dilution was such that it almost
vanishes from the map. We have checked, with the available 
redshifts, that the 2D NE maximum is not due to a superposition
effect, and visual inspection of recent deep CCD imaging
 confirms the reality of this structure (S. Maurogordato, private comm.).

\subsection{Velocities}

We used published redshifts to obtain the velocity
histogram of the NE and SW galaxy density
peaks. Galaxies used to build each histogram were selected within a
circle of radius 3\arcmin~ from each peak. The mean velocity of the
southwestern peak, derived from 64 galaxy redshifts, is 17590~km
s$^{-1}$. The mean velocity of the northeastern peak, derived from 19
galaxy redshifts, is 16907~km s$^{-1}$. The difference between the
mean velocities of the two galaxy density peaks is very small ($\delta
V\approx 700$ km ~s$^{-1}$) compared to the overall velocity
dispersion of the cluster, indicating that both galaxy density peaks
belong to the same structure.

\begin{figure}[]
\includegraphics[height=9.cm,width=8.7cm,angle=0] {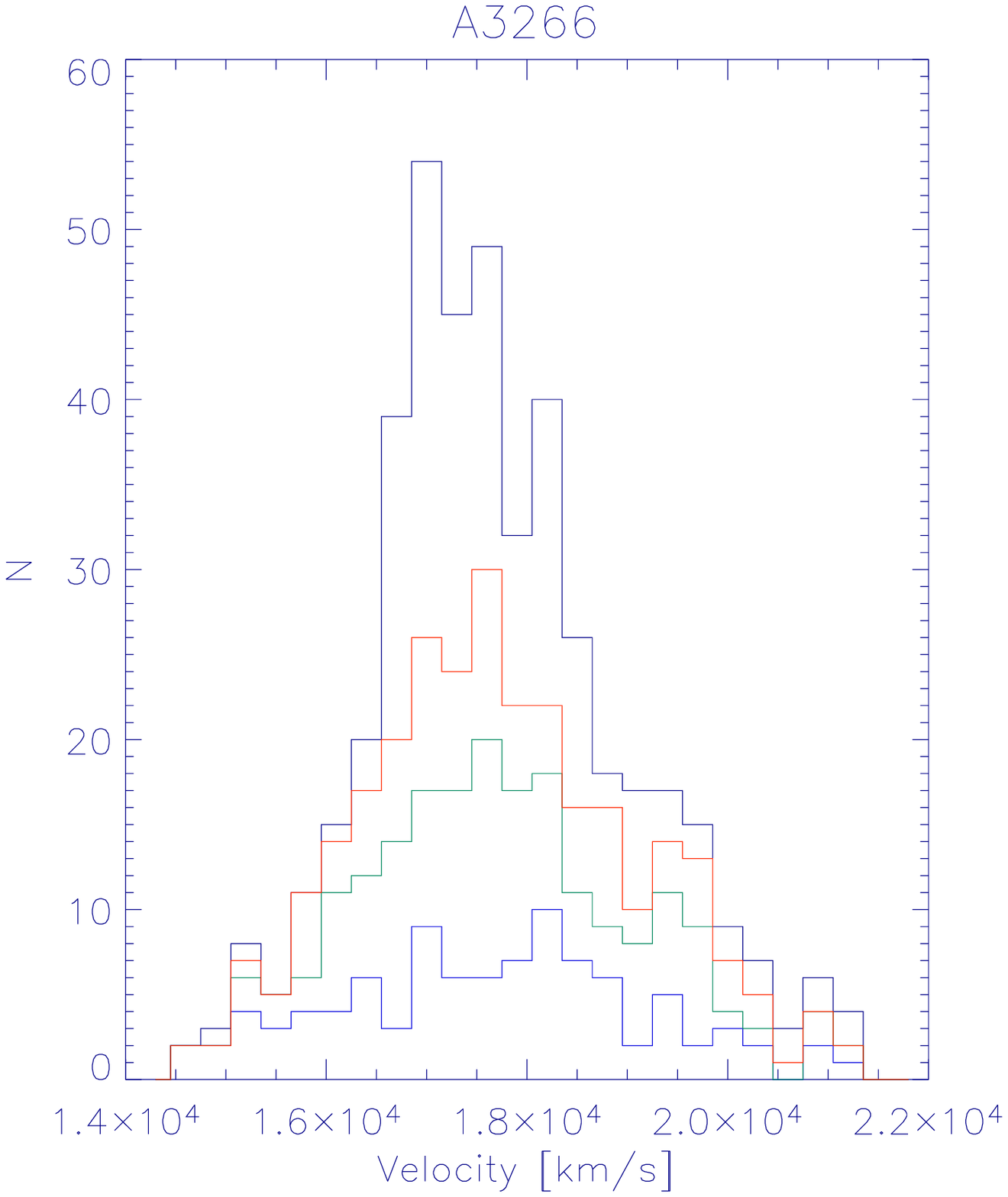} 
\caption{Velocity histograms derived using galaxies
included in circles of increasing radius, from
$R< 10\arcmin$ up to $R< 60 \arcmin$, centred on the BCG 
near the X-ray peak. The mean velocity remains 
stable.}
\label{fig:vel_fctR}
\end{figure}

We used the redshifts published in \citet{qrw96} to analyse the
total velocity distribution in circles of increasing radii centred on
the BCG. We used bins of $300$ km s$^{-1}$ in order to avoid
oversampling the velocity data.  
Figure~\ref{fig:vel_fctR} displays four histograms for circles of maximum 
radius r$_{\rm max}$ of 10,20,30 and 60 arcminutes. 

The galaxy velocity distribution is single-peaked and the dispersion
decreases with increasing encircled radius (see also
\citealt{qrw96}). The latter may be due to a reduction of the
projection effect. The mean velocity changes only slightly out to a
radius of 60\arcmin~ from the centre. We observe a high-velocity tail,
particularly in the histogram of galaxies within $60\arcmin$. As
expected, inspection of the RA-DEC distribution of galaxies with
velocities around 19000 km s$^{-1}$ indicates that they are not
spatially correlated with the NE galaxy density peak shown in
Fig.~\ref{fig:DG}. 

If the galaxy population consisted of two sets of
galaxy over-densities with intrinsically different mean velocities,
such analysis should have detected significant variations in the mean
and shape of the distribution. The above results indicate that the
merger viewing angle is close to the plane of the sky.

\section{Discussion and Tentative scheme}
\label{sec:disc}

\subsection{Viewing angle, merger geometry and mass ratio}

The sharpness of the X-ray temperature (Fig.~\ref{fig:TmapXR}), and
particularly the entropy (Fig.~\ref{fig:PS}) features argue for a
small viewing angle, since projection effects would wash out these
features if the merger was happening at large angles to the line of
sight \citep[e.g.][]{rbl96}.  Furthermore, our optical analysis
indicates that, while there are two peaks in the central ($5\arcmin
\times 5\arcmin$) galaxy density distribution, the velocity
distribution is single peaked from $10\arcmin$ out to $60\arcmin$ from
the centre.  \citet{qrw96} obtained 317 galaxy velocities in a $1\fdg8
\times 1\fdg8$ field, and discussed the possibility of two secondary
peaks in their velocity histogram. However the re-analysis of
\citet{rf00}, using the same data set, concludes that the velocity
distribution is consistent with a single Gaussian, in agreement with
our analysis. \citet{hdd00}, again using the same data set, use the
Kaye's Mixture Model (KMM) algorithm to divide the data into an inner
region and an outer region with slightly different velocity
dispersions. The central velocity of the two Gaussians is not given in
their paper, but close examination of their Fig.~1 shows that they
cannot be statistically different within the (relatively large)
errors.  There is no evidence for a high-velocity component which is
associated with a well-determined galaxy density peak. We thus
conclude that there is no strong evidence for more than one velocity
component from the available optical observations, supporting the
conclusion from the X-ray data that the merger viewing angle is small.

The temperature and entropy features are useful in constraining
the merger geometry and mass ratio. The symmetry of the low energy
X-ray image (Fig.~\ref{fig:DSS_LE}) and temperature map
(Fig.~\ref{fig:TmapXR}) about the NE-SW axis argues
strongly in favour of a small (projected) impact parameter.  On the
other hand, both images are strongly asymmetric along the orthogonal
SE-NW axis.  This asymmetry argues in favour of a merger
event between subclusters of substantially unequal mass.

The observed regularity of the pressure map (right-hand panel of
Fig.~\ref{fig:PS}) can be interpreted in two
ways. Either the present merger has not affected the main cluster as a
whole, and pressure equilibrium still applies more or less everywhere,
or the shock waves have already passed across the entire cluster. In
view of the clear observational evidence of a rather sharp hot
temperature region (Fig.~\ref{fig:TmapXR}), the first option appears
more likely. This argues in favour of a relatively large mass ratio.

We thus continue our discussion in the basic framework of an unequal
mass ratio merger, occurring close-to on-axis at a small to negligible
angle to the line of sight. 

\subsection{Is there a shock ?}

The presence or absence of shocked gas can give key information
on the epoch of the merger event \citep[e.g.][]{mm99}.
Under simple assumptions it is possible to test for shocked gas by
investigating the physical conditions around the edges of the
enhanced temperature region. The inner and outer edges of this region
lie at radii of $\sim 1$\arcmin\ ($68h_{70}^{-1}$ kpc) and
$\sim4$-7\arcmin\ (i.e., 
$270-500h_{70}^{-1}$ kpc), respectively, from the X-ray emission
peak. We apply the 
Rankine-Hugoniot jump conditions for an adiabatic gas: 

\begin{equation}
\frac{1}{C} = \sqrt{4\left(\frac{T_2}{T_1}-1 \right)^2 + \frac{T_2}{T_1}}
-2\left(\frac{T_2}{T_1} -1\right) ~;{\rm for}~ \gamma=\frac{5}{3} ;
\label{eqn3}
\end{equation}
\begin{equation}
\frac{1}{C} =\frac{3}{4}\frac{1}{{\cal M}^2}+\frac{1}{4} ~;{\rm for}~ \gamma=\frac{5}{3},
\label{eqn4}
\end{equation}

\noindent to the region which is assumed to be at the shock
position. We use the measured pre- and post-shock temperatures ($T_1$
and $T_2$, respectively) to
compute the Mach number (${\cal M}$), assuming $\gamma = 5/3$ for an
adiabatic gas. Taking a pre-shock
temperature, estimated from the average of the external western side
of the cluster, of $T_1\simeq7.5$ keV, and a post-shock
temperature, as measured in the enhanced temperature region, of
$T_2\simeq9.5$ keV, we find $\frac{\rho_1}{\rho_2}=\frac{1}{C}=0.71$
and a Mach number ${\cal M}\approx1.3 $.

Noting that the angle defined by the `arms' of the enhanced
temperature region is roughly $\phi\approx120^{\circ}$ leads to an 
alternative calculation of ${\cal M}$. If one interprets this angle as
the Mach cone, then the Mach number should be ${\cal
  M}=\frac{1}{sin(\phi/2)}\simeq1.2$. 

These two independent estimates indicate that, if there is a shock, it
is weak. Further evidence comes from the emissivity and entropy
images. These are both very regular and do not show any significant
gradient perpendicular to the outer temperature edge. This also argues
against the existence of a strong shock here.

\subsection{Merger age}

\begin{figure*}[t!]
\vspace{-0.25cm}
\begin{centering}
\includegraphics[scale=0.55,angle=0,keepaspectratio,width=\textwidth]{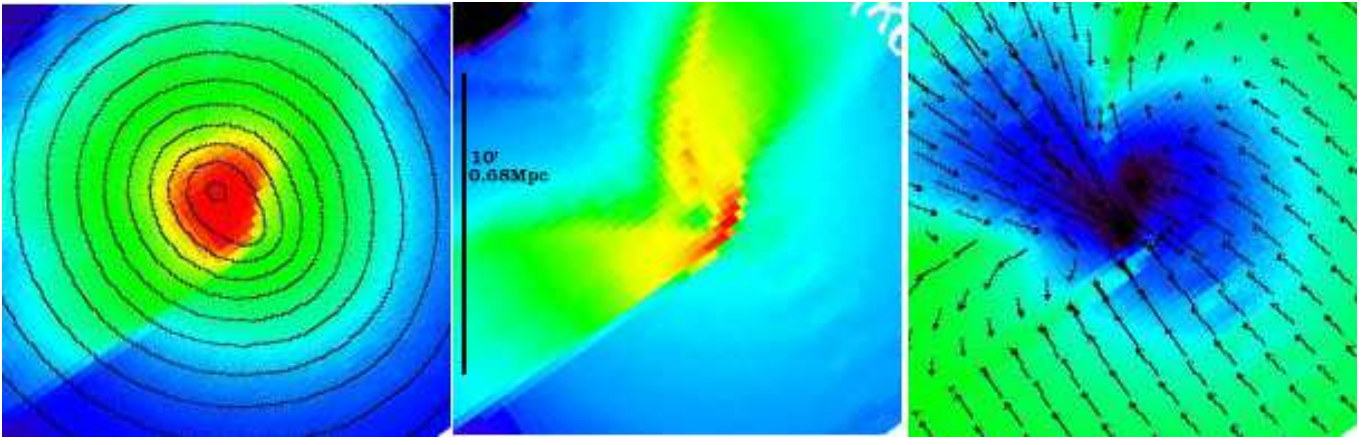}
\caption{{\footnotesize Images from the simulations of a 1:3 mass
ratio, $b=2r_s$ impact parameter merger by \citet{rs01}. In this
scenario, the subcluster has entered the cluster from the northeast
direction. The three
panels show the status of the merger just 0.2 Gyr after the first
(maximum) core collapse.  From left to right: gas density with
contours of gravitational potential, gas temperature (red:hot,
blue:cold), and gas entropy (blue: low entropy, green: high entropy)
with velocity field (arrows). The slight asymmetry is due to the non-zero
impact parameter. The bar in the middle panel is 10\arcmin\ ($\sim
680h_{70}^{-1}$ kpc at the distance of A3266) in length. Each image
has been rotated to match the observed X-ray morphology of A3266.
 }} 
\label{fig:m13_dts}
\end{centering}
\end{figure*}

The new result of the XMM/EPIC observation is the very
structured temperature enhancement surrounding the cool cluster centre
in the SW direction. Since this feature is at the scale of the whole
cluster, it is most likely to be associated with the last merger
event, and not due to one or more older events such as seen in, e.g., A1750
\citep{bel04}. In addition, the brightest galaxy shows a
dumb-bell morphology and these structures are thought to form when the
cores of two galaxy clusters collide \citep{tremaine90,qrw96}. The
structure observed in the temperature and entropy of the ICM, together
with this piece of information, strongly favour the possibility that
the two cluster cores have already been in contact (see also
\citealt{qrw96}, \citealt{fqw00}, \citealt{ht02}). The question is: 
what is the current merger epoch?

Advances in numerical modelling now allow us to
compare our results with realistic, if idealised, simulations of
cluster mergers\footnote{Realistic in the sense that the gas lies in a
Navarro-Frenk-White dark matter halo; idealised in the sense that the
cluster mergers have strictly-defined initial conditions, and do not
take place in a hierarchical cosmological context.}. We used the
simulations of slightly off-axis (impact parameter $b=2r_s$), 1:3 mass
ratio mergers performed by \citet{rs01}\footnote{Available at {\tt
www.astro.uiuc.edu/$\sim$pmricker/research/}} as a comparison with our
X-ray data. We searched these simulations for periods when i) the gas
density was reasonably regular but displayed a small amount of
asymmetry in the centre, ii) the temperature distribution was mainly
characterised by a bow-shaped region of higher temperature, and iii)
the gas entropy showed the comet-tail shape observed in the left and
middle panels of Fig.~\ref{fig:PS}. Two different merger epochs answer
to these criteria. 

\subsubsection{Epoch:  +0.2 Gyr after first core collapse}

Figure ~\ref{fig:m13_dts} shows the density, temperature and entropy
distributions at 0.2 Gyr after the first (maximum) core collapse. 
It can be seen that this merger epoch reproduces the essential
characteristics of the X-ray morphology, temperature and entropy
structure. It is particularly striking how well the observed low
temperature core and surrounding high temperature region are
reproduced. In this scenario 
a small group, with lower temperature and entropy, has entered the main
cluster from the NE and is now exiting in the SW
direction, having 
passed the main cluster core $\sim 0.15-0.20$ Gyr ago. The group gas was
stripped as it entered regions of progressively denser cluster gas,
leaving behind a wake of low-entropy gas. This may explain why the
density gradient seen in the \cha\ data closely follows the entropy
contours. The compression/shock wave would propagate in the direction of
motion of the subcluster (i.e. toward the SW), thus explaining the
temperature distribution. Since the cores of the two clusters have
already been in contact, this collision can also explain the formation
of the dumb-bell galaxy.
 
Interestingly, we can arrive at a similar merger epoch by associating
the two galaxy 
density peaks discussed in Section~\ref{sec:galden} with the
cluster-subcluster pair, and using the ``ballistic'' equations
(e.g. \citealt{sar02}). From Fig.~8 of \citet{rs01}, we can estimate that the
global temperature has increased by a factor $\sim2.2$ due to the
merger. The most massive unit thus had k$T\approx3$ keV before the
merger event, leading to an estimated mass of $\sim 2 \times
10^{14}~M_{\odot}$ using the $M-T$ relation of 
\citet*{momoT}.
The projected distance measured between the two
galaxy density peaks is $\approx5.2$\arcmin\ or $d=0.36$ Mpc in our cosmology.
Assuming that the galaxies follow the dark matter, this distance is
representative of the distance between the two potential wells. The 
turn-around distance is then:

\begin{equation}
d_0\approx 4.5 \left(\frac{M_1+M_2}{10^{15}M_{\odot}}\right)^{1/3}
t_{merge({\rm Gyr})}^{2/3} ~{\rm Mpc}
\label{eqn1}
\end{equation}

\noindent where $t_{merge}$ is the ``absolute'' date of
the merger \citep{sar02} given by:
\begin{equation}
t_{merge}\approx13.7-980 \frac{d_{\rm Mpc}} {v_{\rm km~s^{-1}}} ~{\rm Gyr}, 
\label{eqn0}
\end{equation}
if we consider the age of the Universe equal to 13.7 Gyr in our cosmology.
The merger velocities can be estimated using: 

\begin{equation}
v\approx2930 \left(\frac{M_1+M_2}{10^{15}M_{\odot}}\right)^{1/2}
d_{Mpc}^{-1/2}\left[\frac{1-\frac{d}{d_0}}{1-(\frac{b}{d_0})^2}
\right]^{1/2} {\rm km~s}^{-1},
\label{eqn2}
\end{equation}

\noindent where $d$ is the true distance between the units and $b$ is
the impact parameter. In our case, we use an impact parameter $b=0$,
noting that the result does not strongly depend on the exact value
\citep{sar02}. Assuming 
a mass ratio of $\sim 1:3$, we find $v\approx$2400 km s$^{-1}$, and a
time since maximum collapse of 0.15 Gyr. Notice that we arrived at
this estimate using only the inferred mass ratio, the distance between
the two galaxy density peaks, and the observed average X-ray
temperature.

As outlined above, this scenario requires us to associate the two
galaxy density peaks discussed in Section~\ref{sec:galden} with the
cluster-subcluster pair, where the subcluster, travelling from the NE
towards the SW, has just passed the core of the main cluster. This
interpretation requires us to associate the southwestern galaxy
density peak with the subcluster, and the NE density peak with
the main cluster. We caution that the spectroscopic completeness of
the optical data is not excellent: of the 42 galaxies with $R<19$ in
the APM catalogue in the central $5\arcmin \times 5\arcmin$, only 12
have redshift determinations. Deeper optical observations are required
to probe the very central regions of the cluster.

\subsubsection{Epoch:  +0.8 Gyr after first core collapse} 

\begin{figure*}
\vspace{-0.25cm}
\begin{centering}
\includegraphics[scale=0.55,angle=0,keepaspectratio,width=\textwidth]{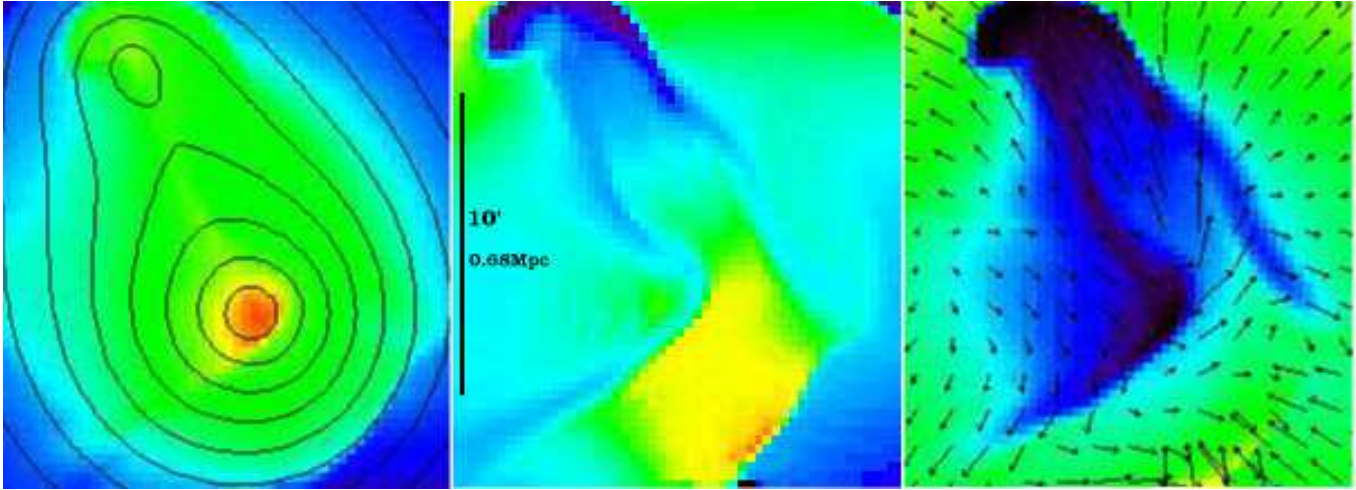}
\caption{{\footnotesize Images from the simulations of a 1:3 mass
    ratio, $b=2r_s$ 
impact parameter merger by \citet{rs01}. In this scenario, the
subcluster has entered the main cluster from the SW
direction, and is approaching turnaround. The three panels show the 
status of the merger 0.8 Gyr after the first (maximum) core
collapse.  From left to right: gas density with contours of
gravitational potential, gas temperature (red:hot,
blue:cold), and gas entropy (blue: low entropy, green: high entropy) with
velocity field (arrows). The slight asymmetry is due to the non-zero
impact parameter. The bar in the middle panel is 10\arcmin\ ($\sim
680h_{70}^{-1}$ kpc at the distance of A3266) in length. Each image
has been rotated to match the observed X-ray morphology of A3266. } }
\label{fig:m13_dts2}
\end{centering}
\end{figure*}

Returning to the simulations of \citet{rs01}, we notice that the main
morphological, temperature and entropy criteria are also approximately
fulfilled at 0.8 Gyr after core passage (Fig.~\ref{fig:m13_dts2}) in
the same off-axis, 1:3 mass ratio simulation. In this scenario, the
subcluster has entered the main cluster from the SW, passing the core
$\sim 0.8$ Gyr ago, exiting towards the NE. At this point in the
simulation, the subcluster is nearing turnaround, and the primary
shock has dissipated far from the main cluster core, well beyond the
edge of the exiting subcluster and outside the range of our X-ray
observations. The heated central region in this case is the
countershock, which is propagating in the opposite direction to the
primary shock. Low entropy, high metallicity gas has been stripped from
the subcluster, leaving behind a low entropy wake with an
inhomogeneous distribution of metals. This explanation was also put
forward by \citet{hdd00} and \citet{ht02}, where they associated a
galaxy concentration $\sim 16\arcmin$ to the NE (i.e., beyond
the field of view of our optical observations) with the exited
subcluster.

While the simulation manages to match the gross X-ray
  characteristics, there are some discrepancies with the
  observations. The overall structure of the hot region in the
  temperature map (Fig.~\ref{fig:TmapXR}), is less well reproduced
  in the simulation (middle panel of Fig.~\ref{fig:m13_dts2}).
  Furthermore, the subcluster is clearly
  visible at $\sim 10\arcmin$ from the main cluster towards the
  NE corner of the simulated gas 
  density map shown in the left panel of Fig.~\ref{fig:m13_dts2}. It is
  also visible as a region of low temperature and entropy in middle
  and right hand panels of Fig.~\ref{fig:m13_dts2}. It is possible
  that the X-ray gas associated with the subcluster was entirely
  stripped during its passage through the main cluster, which may
  imply a more elevated mass ratio than that considered in these
  simulations. This interpretation is bolstered by the lack of X-ray
  emission at the position of the optical subcluster in the {\it
    ROSAT\/} X-ray maps. 

\subsubsection{The wide-angle-tail galaxy}

One further observational aspect is not explained by either of the two
above hypotheses: the orientation of the lobes
of the wide-angle-tail (WAT) galaxy located $\sim7$\arcmin\ to the
SW of the X-ray 
peak. The radio lobes of this galaxy are oriented in the SW-NE
direction of the merger. Among other interpretations, it is suggested that
 radio lobes are bent by bulk motions of the
medium in which the radio galaxy lie in (e.g., \citealt*{hsw05}, and
references therein), 
especially when these galaxies 
are found near the centre of clusters and are expected to have only a small 
relative velocity with respect to the centre of the cluster
itself. This is not the case for the WAT galaxy in A3266, which is i)
found in rather peripheral position and  ii) associated with a host
galaxy of relative 
velocity $\sim800$ km s$^{-1}$ less than the mean cluster
velocity (although this is still within the large dispersion of the
whole cluster). To explain the direction of the bent radio lobes, bulk
gas motion of order 2000 km s$^{-1}$ would be needed. The order of
magnitude is the same as we find from our simple dynamic
calculation in the epoch +0.2 Gyr scenario. However, in this case, the
shock front has not yet had the time to reach the 
location of the WAT. It is even more difficult to explain the
orientation of the WAT lobes in the +0.8 Gyr scenario  without
invoking large proper 
motion of the galaxy itself (which in this case should have belonged
to the smaller cluster) at the time of collision of the two objects.

\section{Conclusion}

We have used a set of 5 XMM observations of the merging 
cluster A3266 to build a mosaic extending to 20 \arcmin~
(i.e. $\approx1.4$ Mpc) from the 
cluster centre. We have been able to derive precise
density, temperature, pseudo-entropy and pseudo-pressure
maps that clearly show a region of enhanced temperature around
a cooler, low entropy core.
The stability of the optical galaxy velocity dispersion argues for a
viewing angle close to the plane of the sky. The symmetry of the low
energy X-ray isophotes and temperature map 
along the proposed collision axis argues for a relatively small impact
parameter. The asymmetry along the axis orthogonal to the collision,
and the relatively undisturbed 
pressure structure, argues for a mass ratio significantly greater than
one. An optical galaxy density map suggests the presence of
substructure in the central region, in the form of two density peaks,
separated by $360 h_{70}^{-1}$ kpc, lying to the NE and SW
 of the X-ray maximum. 

Using close comparison with numerical simulations of unequal mass
ratio, slightly off-axis merger events \citep{rs01}, we have arrived at two
possible interpretations for the observed X-ray and optical structure.
One interpretation suggests that the subcluster entered the main
cluster from the NE, passing the core of the main cluster
0.15-0.20 Gyr, generating the shock wave which is now propagating
ahead of the subcluster toward the outskirts of the newly formed
cluster. A dynamical calculation made by associating the central
galaxy density peaks with the cluster-subcluster pair, leads to a very
similar estimate for the merger epoch. However, completeness issues with
the optical spectroscopic observations do not allow us to make
definite statements concerning the central galaxy density peaks.

An alternative explanation can be found if the subcluster has entered
the main cluster from the SW, passing the main cluster core
some 0.8 Gyr ago, and exiting towards the NE. At this point in the
simulation, the subcluster is near to turnaround. Here the primary
shock is beyond the reach of our X-ray 
maps, in the low-density regions at the edge of the cluster. The
region of heated gas near the core is the countershock, 
propagating in the opposite (SW) direction to the primary
shock. The lack of X-ray emission to the NE
of the X-ray
peak in the larger-scale {\it ROSAT\/} observation of this cluster,
which would be 
associated with the subcluster potential, would suggest
that the subcluster gas has been entirely stripped during the
encounter, implying a larger mass ratio than that considered in the
simulations.

Deeper optical observations are required to give improved constraints
on the relative size of the central galaxy density peaks.  
 
\begin{acknowledgements}
We are grateful to S. Maurogordato for providing the galaxy density
map, and to H. Bourdin for the use of his wavelet based temperature
mapping code. We thank P. Ricker for making his simulations available
on the web, and M.J. Hardcastle for discussions about WAT
galaxies. We acknowledge the Programme National de Cosmologie (PNC)
for supporting the collaboration between the Observatoire de la C\^ote
d'Azur and the Service d'Astrophysique, CEA-Saclay. EB acknowledges
support from PPARC; GWP acknowledges support from a Marie Curie
Intra-European Fellowship under the FP6 programme (contract
no. MEIF-CT-2003-500915). 

The paper is based on observations obtained with \xmm, an ESA science
mission with instruments and contributions directly funded by ESA
Member States and the USA (NASA). This research has made use of: the
XMM-Newton Science Archive (XSA) data base and the \cha\ archive
database, NASA's Astrophysics Data System Abstract Service, the SIMBAD
database operated at CDS, Strasbourg, France, the High Energy
Astrophysics Science Archive Research Center Online Service, provided
by the NASA/Goddard Space Flight Center, and the Digitized Sky Surveys
produced at the Space Telescope Science Institute.

\end{acknowledgements}

\end{document}